\documentclass[pra]{revtex4}
\usepackage{amssymb}
\usepackage{amsmath}

\newcommand{\vc}[1]{\boldsymbol{#1}}

\begin{document}

\title{Constructing Hermitian Hamiltonians for  spin zero neutral and charged particles on a curved surface : physical approach }

\author{M. S. Shikakhwa}
\affiliation{Physics Group, Middle East Technical University Northern Cyprus Campus,\\
Kalkanl\i, G\"{u}zelyurt, via Mersin 10, Turkey}

\author{N.Chair}
\affiliation{Department of Physics,University of Jordan,\\Queen Rania Street,\\
Amman, Jordan}

\begin{abstract}
The surface Hamiltonian for a spin zero particle that is pinned to a surface by letting the thickness of a layer surrounding the surface go to zero - assuming a strong normal force- is constructed. The new approach we follow to achieve this is to start  with an expression for the 3D  momentum operators whose components along the surface and the normal to the surface are separately Hermitian. The  normal part of the  kinetic energy operator is a Hermitian operator in this case. When this operator is dropped and the thickness of the layer is set to zero, one automatically  gets the Hermitian surface Hamiltonian that contains the geometric potential term as expected. Hamiltonians for both a neutral and a charged particle in an electromagnetic field are constructed. We show that a Hermitian surface and normal momenta emerge automatically once one symmetrizes the usual normal and surface momentum operators. The present approach makes it manifest that the geometrical potential originates from the term that is added to the surface momentum operator to render it Hermitian; this term itself emerges from symmetrization/ordering of differential momentum operators in curvilinear coordinates.  We investigate the connection between this approach and the similar approach of Jenssen and Koppe and Costa ( the so called Thin-Layer Quantization (TLQ)). We note that the critical transformation of the wavefunction introduced there before taking the thickness of the layer to zero actually - while not noted explicitly stated by the authors- renders each of the surface and normal kinetic energy operators Hermitian by itself, which is just what our approach does from the onset.
\end{abstract}

\maketitle
\section{Introduction}
The revival of the interest in the quantum mechanics of particles on surfaces and curves in the last decad is evidently due to the advance in technology that made it possible to fabricate nano-scale curved geometries like  nano-spheres, nano-tubes and nano-wires...etc.
A major and well-established  approach for the problem introduced first by Jenssen and Koppe \cite{Koppe} and then elaborated on by Costa \cite{Costa} is the so called thin layer quantization (TLQ). The idea is to first embed the 2D surface in a 3D layer of thickness $d$ and then, by introducing a strong confining potential in the direction normal to the surface to pin the particle to the surface. The part of the Hamiltonian containing the normal degrees of freedom  is then ignored on the ground that the transverse excitations for a sufficiently strong confining potential have a much higher energy than those at the surface, and so can be safely neglected in comparison to the range of energies considered. Mathematically, this amounts to taking the limit $d\rightarrow 0$ "correctly"  in the 3D Schr\"{o}dinger equation/Hamiltonian. "Correctly" here means not to drop a term - called the dangerous term in \cite{Koppe} - that in this limit is actually finite. This  is achieved by implementing a transformation on the wavefunction that isolates the finite term before eventually  taking the limit $d\rightarrow 0$ to obtain the surface Hamiltonian. The finite term in this limit turns out to be a function of the mean and Gaussian curvatures, two geometric invariants of the surface, and is known as the geometric potential or geometric kinetic energy.
Recent works starting from a generalized Dirac quantization approach also obtained the same surface Hamiltonian with the geometric potential \cite{Liu 6}. The TLQ was also applied to construct the surface Hamiltonian of  a spin zero charged particle in an electromagnetic field \cite{ferrari,Jensen1,Jensen2,Ortix1}. Recently, TLQ was also  applied to construct the surface Hamiltonian of a spin one-half particle , especially a one subject to spin-orbit interaction which became a focus of interest by the condensed matter research community \cite{Entin and Magaril,u shape,cheng,exact,Kosugi,SPIN, Wang}. In a recent series of works \cite{shikakhwa and chair1,shikakhwa and chair2,shikakhwa and chair3}, we have developed a variation of the TLQ that was limited, however, to orthogonal curvilinear coordinates. The starting point was still the thin 3D layer and the squeezing potential. The 3D kinetic energy operator, i.e. the Laplacian, was re-written so that the \textit{Hermitian} normal kinetic energy is singled out right from the beginning. The reduction to the 2D surface Hamiltonian was achieved by taking the limit $d\rightarrow 0$ and at the same time dropping this Hermitian normal kinetic energy. The surface Laplacian and the geometric potential emerge immediately after taking this limit without the need for a further transformation.

In the present article, we  substantially elaborate on this last  approach in that we first generalize it to hold for general  curvilinear coordinates not only orthogonal ones. We also present it more systematically and consistently by identifying and starting with the Hermitian momentum operators parallel and normal to the surface right from the beginning. This is done for both  neutral and charged particle in an electromagnetic field.  Moreover, we show that the Hermitian momentum operators normal and parallel to the surface emerge naturally as a result of a symmetrization of the derivatives in each of these operators. As a result, we demonstrate  that the geometric potential that appears in the surface Hamiltonian has its origins in the symmetrization/operator ordering of the momentum operators in curvilinear coordinates. Finally, we compare our approach with  TLQ and see that the key transformation carried out in the TLQ in order to obtain the correct surface Hamiltonian actually leads to a Hamiltonian with the normal and parallel kinetic operators separately Hermitian, which is just what we start with in our approach.


\section { The Hermitian surface and normal momentum operators}

We start by considering a spin zero particle in the 3D space that is  confined to a layer of arbitrary thickness $d$ surrounding a surface $\mathcal{S}$ embedded in the space. The space is  spanned by a special set of  curvilinear coordinates \cite {Koppe,Costa} $\{u_i\},i=1..3$   with $u_1$ and $u_2$ lying on the surface,  $\vc{\hat{u}}_3(u_1,u_2)$ the unit vector normal to the surface and $u_3$  the coordinate along that normal. This way, the position vector of the particle $\vc{R}(u_1,u_2,u_3)$ is given as :
\begin{equation}\label{R 3D}
\vc{R}(u_1,u_2,u_3)= \vc{r}(u_1,u_2)+u_3\vc{\hat{u}}_3
\end{equation}
The vectors $\vc{u}_i \equiv \frac{\partial\vc{R}}{\partial u^i}= \partial_i\vc{R}$ tangent to the coordinates are defined as usual, and for the coordinate system defined in Eq.(\ref{R 3D}) read:
 \begin{eqnarray}\label{tangent vectors}
 \vc{u}_a\equiv \partial_a\vc{R} &=& \partial_a\vc{r}+u_3\partial_a\vc{\hat{u}}_3= \vc{a}_a+u_3\partial_a\vc{\hat{u}}_3\\
  \vc{u}_3 \equiv\partial_3\vc{R} &=&\vc{\hat{u}}_3
 \end{eqnarray}
 where we have defined the tangent vectors on the surface
 \begin{equation}\label{a vectors}
\vc{a}_a\equiv\partial_a\vc{r}
 \end{equation}
and the indices $a,b,c..$ running over $1,2$ refer to the surface degrees of freedom.
In the coordinate system at hand, the metric tensor in the  3D space  $G_{ij}=\partial_i\vc{R}\cdot\partial_j\vc{R}$  assumes the form:
\begin{equation}\label{metric G}
G_{ij}=\left(\begin{array}{cc}
         G_{ab} & 0 \\
         0 & 1
       \end{array}
       \right )
\end{equation}

 $G_{ab}$  can be expressed in terms of the surface metric $g_{ab}\equiv\partial_a\vc{r}\cdot\partial_b\vc{r}=\vc{a}_a\cdot \vc{a}_b $. The two metric tensors are related as \cite{Koppe,Costa}
\begin{eqnarray}\label{G and g}
  G_{ab} &=& (\vc{a}_a+u_3\partial_a\vc{\hat{u}}_3)\cdot (\vc{a}_b+u_3\partial_b\vc{\hat{u}}_3)\\
   &=& g_{ab}-2u_3 K_{ab}+(u_3)^2K_{ac}K_b^c
\end{eqnarray}
where  $K_{ab}=K_{ab}(u_1,u_2)$ is the symmetric curvature tensor of the surface defined as \cite{lecture notes}:
\begin{equation}\label{K definition}
\partial_a\vc{\hat{u}}_3=-K_{ab}\vc{a}^a
\end{equation}
The components of $K_{ab}$ are the  projections of the derivative of the normal along the surface tangent vectors $\vc{a}^a$. The determinant of the metric $G\equiv det\quad G_{ij}= det \quad G_{ab}$  can be calculated from Eq.(\ref{G and g}) and reads  :
 \begin{equation}\label{G}
G (u_1,u_2,u_3)=g(1-4Mu_3+(4M^2+2K)u_3^2+O(u_3^3))
 \end{equation}
 $g=g(u_1,u_2)= det\quad g_{ab}$; $M$ and $K$ are, respectively, the mean and the Gaussian curvatures of the surface defined as:
 \begin{eqnarray}\label{M}
   M &=& \frac{1}{2}K_a^a \\\nonumber
    &=& \frac{g_{22}K_{11}+g_{11}K_{22}-2g_{12}K_{12}}{g}
 \end{eqnarray}
 and,
 \begin{eqnarray}\label{K}
   K &=& det\quad K^a_b \\\nonumber
    &=& \frac{K_{11}K_{22}-(K_{12})^2}{g}
 \end{eqnarray}
 We will also need $\sqrt{G}$ which from Eq.(\ref{G})reads:
 \begin{equation}\label{sqrt G}
 \sqrt{G}= \sqrt{g}\gamma=\sqrt{g}(1-4Mu_3+(4M^2+2K)u_3^2+O(u_3^3))^{\frac{1}{2}}
 \end{equation}

 Note that the $u_3$-dependence in $\sqrt{G}$ lies exclusively in $\gamma$ . To second order in $u_3$ this latter reads :
\begin{equation}\label{gamma}
\gamma=(1-4Mu_3+(4M^2+2K)u_3^2+O(u_3^3))^{\frac{1}{2}}=1-2Mu_3+Ku_3^2+O(u_3^3)
\end{equation}

 We now construct the Hermitian radial and surface momentum operators. The Hermicity of the  3D  momentum operator $ \vc{p}=-i\hbar\nabla$ should be preserved when it is expressed in general curvilinear coordinates where it reads $ \vc{p}=-i\hbar\nabla= -i\hbar\vc{u}^i\partial_i$ (recall that $\vc{u}_i \equiv\partial_i\vc{R}$). Checking, we find :
\begin{equation}\label{Hermicity of p}
\langle \Psi|\vc{p}\Psi\rangle=\langle\Psi| \vc{p}\Psi\rangle +\langle \Psi|\frac{-i\hbar}{\sqrt{G}}\partial_i( \sqrt{G}\vc{u}^i) \Psi\rangle
\end{equation}
where integration is over all space with the measure $\sqrt{G} d^3u$ and the wavefunction was assumed to satisfy boundary conditions that allows the surface term to be dropped . Hermicity of $\vc{p}$ demands the vanishing of the second term on the r.h.s, i.e.:
\begin{equation}\label{identity}
\frac{1}{\sqrt{G}}\partial_i( \sqrt{G}\vc{u}^i)=\frac{1}{\sqrt{G}}\partial_3( \sqrt{G}\vc{\hat{u}}_3)+\frac{1}{\sqrt{G}}\partial_a( \sqrt{G}\vc{u}^a)=0
\end{equation}
where the tangent vectors $\vc{u}^a$ and the normal unit vector $\vc{\hat{u}}_3$ have been defined in Eq.(\ref{tangent vectors}). The above is in fact an identity and can be easily proven. One just needs to note that (see \cite{lecture notes}) $\partial_i\vc{u}^i=\Gamma_{ik}^i\vc{u}^i$ and $\Gamma_{ij}^i=\frac{1}{\sqrt{G}}\partial_j\sqrt{G}$ where $\Gamma ^i_{jk}$ refer to Christoffel Symbol of the second kind. The 3D momentum operators explicitly expressed as a sum of the  surface and normal parts reads:
\begin{eqnarray}\label{P split}
  \vc{p} &=& \vc{p'}+\vc{p}_3 \\\nonumber
 &=&-i\hbar\vc{u}^a\partial_a+ -i\hbar\vc{\hat{u}}_3 \partial_3
\end{eqnarray}
It is straightforward to check  that neither the surface nor the normal momentum as they stand are Hermitian in 3D space; only their sum is. Now adding (half) of the \textit{zero-valued} expression in Eq.(\ref{identity}) multiplied by $-i\hbar$ and split among $\vc{p'}$ and $\vc{p}_3$ to the above $\vc{p}$, we have :
\begin{equation}\label{P split Hermitian}
\vc{p} = \vc{p'}+\vc{p}_3  = \vc{p'}_H+\vc{p}_{3H}
\end{equation}
with the operators $\vc{p'}_H$ and $\vc{p}_{3H}$ being now:
\begin{eqnarray}\label{Hermitian P'}
  \vc{p'}_H &=& \vc{p'}-\frac{i\hbar}{2\sqrt{G}}\partial_a( \sqrt{G}\vc{u}^a)=\vc{p'}+\frac{i\hbar}{2\sqrt{G}}\partial_3( \sqrt{G}\vc{\hat{u}}_3) \\\nonumber
   &=&-i\hbar(\vc{u}^a\partial_a+ \vc{\hat{u}}_3F(G))
\end{eqnarray}
and,
\begin{eqnarray}\label{Hermitian P3}
  \vc{p}_{3H} &=&\vc{p}_3-\frac{i\hbar}{2\sqrt{G}}\partial_3(\sqrt{G}\vc{\hat{u}}_3)  \\\nonumber
   &=& -i\hbar \vc{\hat{u}}_3(\partial_3-F(G))
\end{eqnarray}
and we have defined $F(G)$ as:
\begin{equation}\label{F(G)}
F(G)=-\frac{1}{2\sqrt{G}}\partial_3(\sqrt{G})=M+(2M^2-K)u_3+O(u_3^2)
\end{equation}
and noted that $\partial_3\vc{\hat{u}}_3=0$. The newly defined surface momentum $ \vc{p'}_H$ and normal momentum $\vc{p}_{3H}$ can be readily checked to be Hermitian over the 3D space. Therefore, by adding a zero-valued quantity to the full 3D Hermitian momentum operator, we managed to express it as the sum of Hermitian surface and normal momenta. This  a key step for the following analysis.

\section{Hermitian Hamiltonian for a spin zero neutral particle}

The Hamiltonian for a spin zero particle confined to the  curved surface and otherwise free will now be constructed. The approach is based on the intuitive argument that if one starts  from the full Hamiltonian in the 3D space spanned by the coordinate system given in Eq.(\ref{R 3D}) and then confine the particle to the surface by introducing a strong confining potential (force) along the direction normal to the surface, then the excitation  along this direction will need an infinite energy and so the dynamics is essentially along the surface. This amounts to freezing the  normal degree of freedom  and dropping it from the Hamiltonian which is achieved by setting $d$ to zero and dropping the differential operators with respect to the normal variable $u_3$. The critical  point of the present approach is to drop the\textit{ Hermitian normal momentum}  so that one is left with a Hermitian surface Hamiltonian. While this intuition might seem obvious and trivial, blindly dropping the normal degrees of freedom without observing for Hermicity has led to the reporting of non-Hermitian surface Hamiltonians in the literature \cite{Lyanda-Geller}.Therefore, the essential starting point is  the expression of the 3D momentum operator as a sum of the two Hermitian surface and normal momentum operators as in Eq.(\ref{P split Hermitian}).
 The free particle Hamiltonian in 3D with the momentum operator given by Eq.(\ref{P split Hermitian}) reads :
\begin{equation}\label{3D Hamiltonian}
 H=\frac{p^2}{2m}=\frac{1}{2m}(p_{3H}^2+p_H^{'2}+\vc{p'}_H\cdot \vc{p}_{3H}+\vc{p}_{3H}\cdot\vc{p'}_H)
\end{equation}

Delicate calculation gives the following expressions for each term of the above Hamiltonian:
\begin{equation}\label{P'H dot P3}
\frac{1}{2m}(\vc{p'}_H\cdot \vc{p}_{3H}+\vc{p}_{3H}\cdot\vc{p'}_H)=\frac{-\hbar^2}{2m}\partial_3F(G)
\end{equation}
\begin{equation}\label{P'H^2}
 \frac{p_H^{'2}}{2m}=\frac{-\hbar^2}{2m}(\frac{1}{\sqrt{G}}\partial_a\sqrt{G}G^{ab}\partial_b-F^2(G))
\end{equation}
So, the Hamiltonian, Eq.(\ref{3D Hamiltonian}), becomes:
\begin{equation}\label{3D Hamiltonian expanded}
H=\frac{p_{3H}^2}{2m}+\frac{-\hbar^2}{2m}(\frac{1}{\sqrt{G}}\partial_a\sqrt{G}G^{ab}\partial_b+\partial_3F(G)-F^2(G))
\end{equation}
The last two terms  in the bracket on the r.h.s. of the above expression for the Hamiltonians are generated by the extra terms that were added to $\vc{p'}_H$ and $\vc{p}_{3H}$ to render each Hermitian. Evidently, if one expands $\frac{p_{3H}^2}{2m}$  - which we will not do !- identical terms with opposite signs will be emerge leading to full cancellation of these terms and the usual expression for the 3D free particle Hamiltonian will be restored.  Now, using the explicit expressions of $G_{ab}$ and $F(G)$ given, respectively, by Eqs.(\ref{G and g}) and (\ref{F(G)})we can easily verify that as $u_3\rightarrow 0$:
\begin{eqnarray}\label{u3 = 0 limits 1}
 \frac{1}{\sqrt{G}}\partial_a\sqrt{G}G^{ab}\partial_b |_{u_3\rightarrow 0}&\rightarrow& \frac{1}{\sqrt{g}}\partial_a\sqrt{g}g^{ab}\partial_b\\
  F^2(G) |_{u_3\rightarrow 0}&\rightarrow& M^2 \\\label{u3 = 0 limits 2}
  \partial_3F(G)|_{u_3\rightarrow 0}&\rightarrow& 2M^2-K\label{u3 = 0 limits 3}
\end{eqnarray}
Therefore, the dropping of the normal degree of freedom from the Hamiltonian, Eq.(\ref{3D Hamiltonian expanded}), will be achieved by dropping the normal Hermitian momentum operator $p_{3H}^2$ and taking the limit $u_3\rightarrow 0$, i.e. employing Eqs.(\ref{u3 = 0 limits 1})-(\ref{u3 = 0 limits 3}). The resulting Hermitian surface Hamiltonian is :
 \begin{equation}\label{H surface}
H_{s}=\frac{-\hbar^2}{2m}(\frac{1}{\sqrt{g}}\partial_a\sqrt{g}g^{ab}\partial_b)-\frac{\hbar^2}{2m}(M^2-K)
 \end{equation}
 This is the well-known expression of the surface Hamiltonian of the TLQ derived first by Costa \cite{Costa}. The first term is the Laplace-Beltrami operator at the surface and the last two terms are the geometric potential or the geometric kinetic energy.

\section{Hamiltonian for a charged spin zero particle in an electromagnetic field}

The surface Hamiltonian for a spin zero charged particle coupled to an electromagnetic field can be constructed along exactly the same lines as that of the neutral particles. A bit of extra care needs to be taken in the calculations. In 3D general curvilinear coordinates, the Hamiltonian for a spin zero particle of charge $q$ coupled to a scalar potential $V(\vc{u})$ and a vector potential $\vc{A(\vc{u})}$ expressed in terms of the kinematic momentum $\vc{\Pi}=(\vc{p}-q\vc{A(\vc{u})})$ reads:
\begin{equation}\label{H em}
H=\frac{\Pi^2}{2m}+qV
\end{equation}
In the  coordinates at hand the vector potential is given as:
\begin{equation}\label{A}
\vc{u}^iA_i=\vc{u}^aA_a+\vc{\hat{u}}^3A_3
\end{equation}
It is crucial here to note that the operator $\vc{p}$ appearing in $\vc{\Pi}$ is ( see Eqs. (\ref{P split Hermitian})-(\ref{Hermitian P3})) the sum of the Hermitian normal and surface momenta. Therefore, the normal and surface components of $\vc{\Pi}$ are given explicitly as:
\begin{eqnarray}\label{Hermitian Pi}
 \vc{\Pi}_{3H}  &=& \vc{p}_{3H}-q\vc{\hat{u}}^3 A_3 = -i\hbar \vc{\hat{u}}_3(\partial_3-F(G)-\frac{iq}{\hbar}A_3) \\
 \vc{\Pi}_{H}  &=& \vc{p'}_{H}-q\vc{u}^aA_a= -i\hbar(\vc{u}^a(\partial_a-\frac{iq}{\hbar}A_a)+\vc{\hat{u}}_3 F(G))
\end{eqnarray}
Thus, the Hamiltonian, Eq.(\ref{H em}), becomes
\begin{equation}\label{H em split}
H=\frac{1}{2m}(\vc{\Pi}_{3H}+\vc{\Pi'}_{H})^2+qV
\end{equation}
The various terms in this Hamiltonian can be calculated in a straightforward manner just as in the spin zero case and read:
\begin{equation}\label{Pi H dot Pi3}
\frac{1}{2m}(\vc{\Pi'}_H\cdot \vc{\Pi}_{3H}+\vc{\Pi}_{3H}\cdot\vc{\Pi'}_H)=\frac{-\hbar^2}{2m}\partial_3F(G)
\end{equation}
which is the same expression as in the corresponding term in the spin zero case, Eq.(\ref{P'H dot P3}). Also,
\begin{equation}\label{Pi'H^2}
 \frac{\Pi_H^{'2}}{2m}=\frac{p_H^{'2}}{2m}+\frac{i\hbar q}{2m}(\frac{1}{\sqrt{G}}\partial_b(\sqrt{G}G^{ab}A_a))+\frac{i\hbar q}{m}G^{ab}A_a\partial_b
 +\frac{q^2}{2m}G^{ab}A_aA_b
\end{equation}
Substituting for $p_H^{'2}$ from Eq.(\ref{P'H^2}) and expanding the second term in Eq.(\ref{Pi'H^2}) above, we have the explicit form of the Hamiltonian, Eq.(\ref{H em split}):
\begin{eqnarray}\label{H em expanded}
  H &=& \frac{\Pi_{3H}^{2}}{2m}-\frac{\hbar^2}{2m}\partial_3F(G)-(\frac{\hbar^2}{2m})\frac{1}{\sqrt{G}}\partial_a\sqrt{G}G^{ab}\partial_b+\frac{\hbar^2}{2m}F^2(G)\\\nonumber
   &+& \frac{i\hbar q}{2m}\left((\partial_b G^{ab})A_a+\frac{1}{\sqrt{G}}(\partial_b(\sqrt{G})G^{ab}A_a+G^{ab}(\partial_b A_a)\right)+\frac{i\hbar q}{m}G^{ab}A_a\partial_b+\frac{q^2}{2m}G^{ab}A_aA_b+qV
\end{eqnarray}

In the limit $u_3\rightarrow 0$, in addition to Eqs,(\ref{u3 = 0 limits 1})-(\ref{u3 = 0 limits 3}) we can easily verify the following :
\begin{eqnarray}\label{u3 = 0 limits 4}
  (\partial_b G^{ab})A_a(u_1,u_2,u_3)|_{u_3\rightarrow 0} &\rightarrow& (\partial_b g^{ab})A_a(u_1,u_2) \\\label{u3 = 0 limits 5}
  \frac{1}{\sqrt{G}}(\partial_b(\sqrt{G})G^{ab}A_a(u_1,u_2,u_3)|_{u_3\rightarrow 0} &\rightarrow& \frac{1}{\sqrt{g}}(\partial_b(\sqrt{g})g^{ab}A_a(u_1,u_2) \\\label{u3 = 0 limits 6}
  G^{ab}(\partial_b A_a(u_1,u_2,u_3))|_{u_3\rightarrow 0} &\rightarrow& g^{ab}(\partial_b A_a(u_1,u_2,u_3)|_{u_3=0}) \\\label{u3 = 0 limits 7}
  G^{ab}A_a(u_1,u_2,u_3)\partial_b|_{u_3\rightarrow 0} &\rightarrow& g^{ab}A_a(u_1,u_2)\partial_b\label{u3 = 0 limits 8}
\end{eqnarray}

At this point, we carry a gauge transformation that eliminates the normal $A_3(u_1,u_2,u_3)$ component of the vector potential. It is always possible to find such a gauge transformation \cite{Weinberg}. The consequence of this is :
\begin{equation}\label{GT of Pi_3}
\vc{\Pi}_{3H}  = \vc{p}_{3H}-q\vc{\hat{u}}^3 A_3\rightarrow  \vc{p}_{3H}=-i\hbar \vc{\hat{u}}_3(\partial_3-F(G))
\end{equation}
which is just the expression of the Hermitian normal momentum in the absence of coupling to the electromagnetic field, Eq.(\ref{Hermitian P3}).
The Hamiltonian, Eq.(\ref{H em expanded}, upon setting $u_3$ and $p_{3H}$  to zero and using the sets of equations (\ref{u3 = 0 limits 1})-(\ref{u3 = 0 limits 3}) and (\ref{u3 = 0 limits 4})-(\ref{u3 = 0 limits 8}))reduces to
\begin{equation}\label{H surface em }
 H|_{u_3\rightarrow 0,P_{3H}\rightarrow 0} \equiv H_s^{em}=-\frac{\hbar^2}{2m}\nabla'^2+\frac{i\hbar q}{2m} \vc{\nabla'}\cdot\vc{A'}(u_1,u_2,u_3)|_{u_3=0}+\frac{i\hbar q}{m}\vc{A'}(u_1,u_2)\cdot\vc{\nabla'}-\frac{\hbar^2}{2m}(M^2-K)+qV+\frac{q^2}{2m}|\vc{A}(u_1,u_2)|^2
\end{equation}
where we have used the surface divergence notation $\vc{\nabla'}\cdot\vc{A'}(u_1,u_2,u_3)|_{u_3=0}\equiv \vc{a}^b\partial_b\cdot A_a\vc{a}^a(u_1,u_2,u_3)|_{u_3=0}$ to denote $\frac{1}{\sqrt{g}}(\partial_b(\sqrt{g}g^{ab}A_a(u_1,u_2,u_3)|_{u_3=0})$, with $\vc{\nabla'}\equiv \vc{a}^b\partial_b$ ; $\vc{A'}\equiv A_b\vc{a}^b$ and $\vc{a}^b=g^{bc}\vc{a}_c$. We have also used the notation $g^{ab}A_a\partial_b=\vc{A'}\cdot\vc{\nabla'}$ and $\nabla'^2$ denotes the surface Laplacian. i.e. the Laplace-Beltrami operator, Eq.(\ref{u3 = 0 limits 1}). The Hamiltobian, Eq.(\ref{H surface em }), was first derived by Ferrari and Coughi \cite{ferrari}.
Finally, we note that we can express this surface Hamiltonian compactly as usual in terms of a covariant derivative   $\vc{D'}\equiv (\vc{\nabla'}-\frac{iq}{\hbar}\vc{A'}(u_1,u_2,u_3))$ as:
\begin{eqnarray}\label{H surface em covariant}
   H_s^{em} &=& -\frac{\hbar^2}{2m}(\vc{\nabla'}-\frac{iq}{\hbar}\vc{A'}(u_1,u_2,u_3))\cdot(\vc{\nabla'}-\frac{iq}{\hbar}\vc{A'}(u_1,u_2,u_3))|_{u_3=0}-\frac{\hbar^2}{2m}(M^2-K)+qV \\\nonumber
   &=& -\frac{\hbar^2}{2m}\vc{D'}\cdot\vc{D'}|_{u_3=0}-\frac{\hbar^2}{2m}(M^2-K)+qV
\end{eqnarray}

\section{Symmetrization of momentum operators, Hermicity and origin of the geometric potential }

Investigating Eqs.(\ref{P'H dot P3})-(\ref{3D Hamiltonian expanded}) and (\ref{H surface}), noting Eqs.(\ref{u3 = 0 limits 1})-(\ref{u3 = 0 limits 3}),we see that the geometric potential $-\frac{\hbar^2}{2m}(M^2-K)$ has its origin in the $F(G)$ term that was introduced into $\vc{p'}$ and $\vc{p_3}$ to render them Hermitian. We now look more closely into this term and try to clarify its meaning. Consider $\vc{p'}_H$ given in Eq.(\ref{Hermitian P'}) and note that it can be symmetrized in $\partial_a$ :
\begin{eqnarray}\label{symmetric P'_H}
   \vc{p'}_H &=& -i\hbar(\vc{u}^a\partial_a+\frac{1}{2\sqrt{G}}\partial_a( \sqrt{G}\vc{u}^a))   \\\nonumber
   &=& -i\hbar(\vc{u}^a\partial_a+\frac{1}{2\sqrt{G}}\partial_a\sqrt{G}\vc{u}^a-\frac{1}{2\sqrt{G}}\sqrt{G}\vc{u}^a\partial_a) \\\nonumber
  &=&\frac{-i\hbar}{2\sqrt{G}}(\sqrt{G}\vc{u}^a\partial_a+\partial_a\sqrt{G}\vc{u}^a)=\frac{-i\hbar}{2\sqrt{G}}\{\sqrt{G}\vc{u}^a,\partial_a\}_{+}
\end{eqnarray}
where in the expression  $\partial_a\sqrt{G}\vc{u}^a$ the derivative acts to everything to its right, and $\{,\}_{+}$ denotes the anti-commutator. So, the term added to render $\vc{p'}$ Hermitian is nothing but a recipe for symmetrizing the derivatives in the specific manner given in Eq.(\ref{symmetric P'_H})in order to construct the Hermitian operator $\vc{p'}_H$. This means that the geometric potential is the outcome of symmetrization or ordering of the derivatives in the momentum operators. It is easy  to symmetrize $\vc{p}_1$ and $\vc{p}_2$ separately and construct the Hermitian version for each of them in the same manner \cite{shikakhwa physica e}. $\vc{p}_{3H}$ can be symmetrized in exactly the same way:
\begin{eqnarray}\label{P3H symmetrized}
  \vc{p}_{3H} &=&-i\hbar\vc{\hat{u}}_3(\partial_3+\frac{1}{2\sqrt{G}}\partial_3(\sqrt{G})) \\\nonumber
   &=&\frac{-i\hbar}{2\sqrt{G}}(\vc{\hat{u}}_3\sqrt{G}\partial_3+\partial_3\sqrt{G}\vc{\hat{u}}_3)=\frac{-i\hbar}{2\sqrt{G}}\{\sqrt{G}\vc{\hat{u}}_3,\partial_3\}_{+}
\end{eqnarray}
Using the above symmetrized expressions for these operators, we can write the 3D momentum operator in a novel symmetrized form as:

  \begin{equation}\label{3D P symmetrized}
 \vc{p}=\vc{p'}_H+\vc{p}_{3H}=\frac{-i\hbar}{2\sqrt{G}}(\{\sqrt{G}\vc{u}^a,\partial_a\}_{+}+\{\sqrt{G}\vc{\hat{u}}_3,\partial_3\}_{+})
\end{equation}
Such symmetrized forms of the momentum can be a good starting point for further calculations sometimes ( see the last paragraph below).
 the limit $u_3\rightarrow 0$ it is easy to check that the symmetrized expressions for the momentum operators reduce to:
\begin{equation}\label{P_3H symm u_3 =0}
 \vc{p}_{3H}=\frac{-i\hbar}{2\gamma}\{\gamma\vc{\hat{u}}_3,\partial_3\}_{+}|_{u_3=0}
\end{equation}
\begin{equation}\label{P_3H symm u3 =0}
  \vc{p'}_H|_{u_3\rightarrow 0}=\frac{-i\hbar}{2\sqrt{g}}\{\sqrt{g}\vc{u}^a,\partial_a\}_{+}
\end{equation}
This last expression for $\vc{p'}_H$ is the symmetric version of what is some times called the geometric momentum at the surface in the literature \cite{Liu2}. One can express the surface Hamiltonian, Eq,(\ref{H surface}), employing Eq.(\ref{P'H^2}) in the limit $u_3\rightarrow 0$ in terms of this momentum to read:
  \begin{equation}\label{H surface P_3H}
H_{s}=\frac{p_H^{'2}}{2m}-\frac{\hbar^2}{2m}(2M^2-K)
 \end{equation}
 The above form of the surface Hamiltonian, with $\frac{p_H^{'2}}{2m}$ symmetrized can be quite convenient in some derivations. In \cite{shikakhwa physica e} it was used to derive the centripetal force on a spin zero particle on the surface of a sphere and a cylinder. Keeping track of the symmetrization all the way, we obtained a symmetrized expression of this force.

\section{Connection with the thin layer quantization }

Here we compare and contrast our approach with the TLQ procedure\cite{Koppe,Costa}. In the TLQ, one also starts with the Schr\"{o}dinger equation  for a particle in a 3D layer of thickness $d$ surrounding a surface $\mathcal{S}$, with the space spanned by the special coordinate system given by Eq.(\ref{R 3D}). Eventually the limit $d\rightarrow 0$ is taken and the Schr\"{o}dinger equation is separated into two equations for the normal and surface degrees of freedom. The key point in the TLQ is a transformation of the wavefunction that isolates a finite term that would otherwise be mistakenly dropped from the Hamiltonian once $d$ is set to zero. The authors of \cite{ Koppe} call this the dangerous term. It turns out that in the limit $d\rightarrow 0$ this finite term is the geometrical potential that was discussed earlier. To be explicit, consider Eq.(14) in the pioneering work \cite{Koppe} which when formulated in the notation used in the present work reads:
\begin{equation}\label{Eq 14}
-\frac{\hbar ^2}{2m}(G^{-\frac{1}{2}}\partial_aG^{ab}G^{\frac{1}{2}}\partial_b\psi+G^{-\frac{1}{2}}\partial_3G^{\frac{1}{2}}\partial_3\psi)=E\psi
\end{equation}
The expression in the brackets on the l.h.s is just the 3D Laplacian split into surface part; the first term, and  the normal part; the second term. The wavefunction $\psi$ is normalized in 3D space with the measure $d^3u\sqrt{G}$ as $\int d^3u\sqrt{G} |\psi|^2=1$. Note that non of the two parts of the Laplacian is Hermitian by itself  , only their sum is. When the transformation
\begin{equation}\label{chi transformation}
\psi=\gamma ^{-\frac{1}{2}}\chi
\end{equation}

is invoked ($\gamma$ defined in \cite{Koppe} is the square of $\gamma$ defined in Eq.(\ref{gamma}) in the present work) , the normal part in the l.h.s of Eq.(\ref{Eq 14}) becomes $\gamma^{-\frac{1}{2}}(\partial_3^2\chi+U\chi)$, with $U$ being the sought for finite term which does not vanish in the limit $d\rightarrow 0$. Note that $\int d^3u\sqrt{G} |\psi|^2=1$ demands that $\int d^3u\sqrt{g} |\chi|^2=1$. The transformed Schr\"{o}dinger equation for $\chi$ is now \cite{Koppe}:
\begin{equation}\label{Eq 14'}
-\frac{\hbar ^2}{2m}(\gamma^{\frac{1}{2}}g^{-\frac{1}{2}}\partial_ag^{ab}g^{\frac{1}{2}}\partial_b\gamma^{-\frac{1}{2}}\chi+\partial_3^2\chi+U\chi)=E\chi
\end{equation}
The differential operator $\partial_3^2$ of the normal variable  can be easily checked to be Hermitian with the measure $d^3u\sqrt{g}$. The surface differential operator, i.e. the first term in Eq.(\ref{Eq 14'}) should also be so since the whole l.h.s. should be Hermitian. Therefore, the transformation, Eq.(\ref{chi transformation}), serves to render each of  the differential operators with respect to the normal and surface parts in the Hamiltonian Hermitian by itself. This is the connection between the TLQ and our approach presented in this work. It is all about having differential  operators for the normal and surface degrees of freedom that are separately Hermitian before setting $d$ to zero. We do this right from the start whereas in TLQ, it takes one to carry out the transformation in Eq.(\ref{chi transformation}) to achieve this. TLQ proceeds by splitting the wavefunction as $\chi=\chi_s(u_1,u_2)\chi_3(u_3)$  so that the Schr\"{o}dinger equation separates into normal and surface equations.   In the limit $d\rightarrow 0$ the first term on the l.h.s of Eq.(\ref{Eq 14'}) reduces to the surface Laplace-Beltrami operator and the potential $U$ to the geometric potential thus one gets the same expression for the surface Hamiltonian as the one obtained by the present approach, Eq.(\ref{H surface}).

\section{Summary and conclusions}

We suggest a new physics-based idea to construct the surface Hamiltonian for a spin zero particle within a layer of thickness $d$ around a surface when it is pinned to the surface as the thickness of the layer is shrunk to zero. The idea is based on starting right from the beginning with the 3D momentum operator $\vc{p}$ (which is the Laplacian in general curvilinear coordinates) that has each of its component operators parallel and normal to the surface Hermitian by itself. This is shown to be achieved naturally by symmetrizing the derivatives in these directions as in Eqs.(\ref{symmetric P'_H}) and (\ref{P3H symmetrized}). This symmetrization is shown to merely adding a zero-valued quantity to $\vc{p}$ so that it is effectively the same. The kinetic energy operators $\frac{p^{2}}{2m}$ has its normal and surface parts Hermitian, too. The surface Hamiltonian with the geometric potential term is obtained by taking the limit $d\rightarrow 0$ and simultaneously dropping the normal kinetic energy operator. The result is the same as that given by the TLQ \cite{Jenssen and Koppe,Costa}. The present approach, however, makes it manifest that the geometric potential that appears in the surface Hamiltonian originates from the symmetrization of the momenta  and the ordering of their differential operators. We have also demonstrated that the manipulations in the TLQ leading to the construction of the surface actually - though not noted explicitly by the authors- serve to render both parts of the kinetic energy operator; the normal and the parallel to the surface, Hermitian by itself before setting the thickness of the layer to zero;  Hermicity, which is the physical criterion observed by our approach is also at the center of TLQ.


\begin{thebibliography}{99}


\bibitem{Koppe} H. Jensen and H. Koppe, Ann. Phys. \textbf{63}, 586(1971)
\bibitem{Costa} R. C. T. da Costa, Phys. Rev. A \textbf{23}, 1982 (1981).
\bibitem {Liu 6} Dingkun Lian,  Liangdong Hu and  Quanhui Liu, Ann. Phys. (Berlin) 1700415 (2018).
\bibitem{ferrari} G. Ferrari and G.Coughi, Phys. Rev. Lett. \textbf{100}, 240403 (2008).
\bibitem {Jensen1} B. Jensen and R. Dandoloff, Phys. Rev. A \textbf{80}, 052109 (2009).
\bibitem{Jensen2} B. Jensen and R. Dandoloff, Phys. Rev. A \textbf{81}, 049905(E) (2010).
\bibitem {Ortix1} Carmine Ortix and Jeroen van den Brink, Phys. Rev. B\textbf{ 83}. 113406 (2011).
\bibitem{Entin and Magaril} M. V. Entin, L. I. Magarill, Phys. Rev. B\textbf{64 }, 085330 (2001).
\bibitem{u shape} M-H. Liu et. al., Phys. Rev. B \textbf{84}, 085307 (2011).
\bibitem{cheng} T-C. Cheng, J-Y. Chen, and C-R. Chang, Phys. Rev. B \textbf{84}, 214423 (2011).
\bibitem{exact} J-Y. Chang, J-S. Wu, and C-R. Chang Phys. Rev. B \textbf{87 }, 174413 (2013).
\bibitem {Kosugi} T.Kosugi, J. Phys. Soc. Jpn. \textbf{80}, 073602 (2011).
\bibitem{SPIN} K-C Chen and C-R. Chang, SPIN \textbf{03 }, 1340006 (2013).
\bibitem{Wang} Y.-L. Wang, H. Jiang and H.-S. Zong, Phys. Rev. A \textbf{96}, 022116 (2017)).
\bibitem {shikakhwa and chair1} M.S.Shikakhwa and N.Chair, Phys.Lett.A \textbf{380},1985 (2016).
\bibitem {shikakhwa and chair2} M.S.Shikakhwa and N.Chair, Phys.Lett.A \textbf{380},2876 (2016).
\bibitem{shikakhwa and chair3}M.S.Shikakhwa and N.Chair, Eur.J.Phys., \textbf{38},015402 (2017).
\bibitem {lecture notes} PA.Kelly, Mechanics Lecture Notes: An introduction to Solid Mechanics.  Available from http://homepages.engineering.auckland.ac.nz/~pkel015/SolidMechanicsBooks/index.html.
\bibitem{Lyanda-Geller} A.G. Aronov, Y.B. Lyanda-Geller, Phys. Rev. B, \textbf{70}, 343 (1993).
\bibitem{Weinberg}  S.Weinberg, The Quantum Theory of Fields Vol.II, Cambridge: Cambridge University Press (1996).
\bibitem{shikakhwa physica e} M.S.Shikakhwa, Physica E: Low-dimensional Systems and Nanostructures, \textbf{108}, 249-252 (2019).
\bibitem {Liu2}Q. H. Liu, C. L. Tong, and M. M. Lai, J. Phys. A \textbf{40},4161 (2007)
















\end{thebibliography}
\end{document}